*Article*

# SCADA System Testbed for Cybersecurity Research Using Machine Learning Approach

**Marcio Andrey Teixeira [1,3]\*, Tara Salman [3], Maede Zolanvari [3], Raj Jain [3], Nader Meskin [2], and Mohammed Samaka [2]**

[1] Federal Institute of Education, Science, and Technology of Sao Paulo, Department of Informatics, Catanduva, SP, Brazil; marcio.andrey@ifsp.edu.br
[2] Qatar University, Doha, Qatar; {samaka.m@qu.edu.qa, Nader.Meskin@qu.edu.qa}
[3] Washington University in Saint Louis, Department of Computer Science and Engineering, Saint Louis, MO, United States; {tara.salman@wustl.edu, maede.zolanvari@wustl.edu, jain@wustl.edu}
\* Correspondence: marcio.andrey@ifsp.edu.br; Tel.: +1 314-201-7439

**Abstract**: This paper presents the development of a SCADA system testbed used for cybersecurity research. The testbed consists of a water storage tank's control system, which is a stage in the process of water treatment and distribution. Sophisticated cyber-attacks are conducted against the testbed. During the attacks, the network traffic is captured, and features are extracted from the traffic to build a dataset for training and testing different machine learning algorithms. Five traditional machine learning algorithms are trained to detect the attacks: Random Forest, Decision Tree, Logistic Regression, Naïve Bayes and KNN. After that, the trained machine learning models are built and deployed in the network, where new tests are made using online network traffic. The performance obtained during the training and test of the machine learning models is compared to the performance obtained during the online deployment of these models in the network. The results show the efficiency of the machine learning models in detecting the attacks in real time. The testbed provides a good understanding of the effects and consequences of attacks on real SCADA environments.

## 1. Introduction

Supervisory Control and Data Acquisition (SCADA) systems are Industrial Control Systems (ICS) widely used by industries to monitor and control different processes such as oil and gas pipelines, water distribution, electrical power grids, etc. These systems provide automated control and remote monitoring of services being used in daily life. For example, states and municipalities use SCADA systems to monitor and regulate water levels in reservoirs, pipe pressure, and water distribution.

A typical SCADA system includes components like computer workstations, Human Machine Interface (HMI), Programmable Logic Controllers (PLCs), sensors, and actuators [1]. Historically, these systems had private and dedicated networks. However, due to the wide-range deployment of remote management, open IP networks (e.g., Internet) are now used for SCADA systems communications [2]. This exposes SCADA systems to the cyberspace and makes them vulnerable to cyber attacks using the Internet.

Machine Learning (ML) and artificial intelligence techniques have been widely used to constitute an intelligent and efficient Intrusion Detection System (IDS) dedicated to ICS. However, researchers generally develop and train their ML-based security system using network traces obtained from publicly available datasets. Due to malware evolution and changes in the attack strategies, these datasets fail to protect the system from the new types of attacks, and consequently, the benchmark datasets should be updated periodically.

This paper presents the deployment of a SCADA system testbed for cybersecurity research and investigates the feasibility of using ML algorithms to detect cyber-attacks in real time. The testbed is built using equipment deployed in real industrial settings. Sophisticated attacks are conducted on the testbed to have a better understanding of the attacks and their consequences in SCADA environments. The network traffic is captured including both abnormal and normal traffic. The behavior of both types of traffic (abnormal and normal) is analyzed, and features are extracted to build a new SCADA-IDS dataset. This dataset is then used for training and testing ML models which are further deployed in the network. The performance of the ML model depends highly on the available datasets. One of the main contributions of this paper is building a new dataset updated with recent and more sophisticated attacks. We argue that IDS using ML models trained with a dataset generated at the process control level could be more efficient, less complicated, and more cost-effective as compared to traditional protection techniques. Five traditional machine learning algorithms are trained to detect the attacks: Random Forest, Decision Tree, Logistic Regression, Naïve Bayes and KNN. Once trained and tested, the ML models are deployed in the network, where real network traffic is used to analyze the effectiveness and efficiency of the ML models in a real-time environment. We compare the performance obtained during the training and test phase of the ML models with the performance obtained during the online deployment of these models in the network. The online deployment is another contribution of this paper since most of the published papers present the performance of the ML models obtained during the training and test phases. We are conducting this research to build an IDS software based on ML models to be deployed on the ICS/SCADA systems.

The remainder of the paper is organized as follows. Section 2 presents a brief background of the ICS-SCADA system reference model and the related works. Section 3 describes the developed SCADA system testbed. Section 4 describes the ML algorithms and the performance measurements used in this work. Section 5 shows the scenario of the conducted the attacks and the main features of the dataset used to train the algorithms. Section 6 discusses our results and the interoperations behind them. Finally, Section 7 concludes the paper with the main discussed points and outcomes.

## 2. Background

In this section, we briefly present a description of the ICS-SCADA reference model and some related works in the domain of ML algorithms for SCADA system security.

*2.1. ICS Reference Model*

The ICS is a general term that covers numerous control systems, including SCADA systems, distributed control systems, and other control system configurations [3]. An ICS consists of combinations of control components (e.g., electrical, mechanical, hydraulic, pneumatic) that are used to achieve various industrial objectives (e.g., manufacturing, transportation of matter or energy). Figure 1 shows an example of an ICS reference model [4].

As can be seen from Figure 1, the ICS model is divided into four levels, 3 to 0. Level 3 (corporate network) consists of the traditional information technology, including the general deployment of services and systems, such as file transfer, websites, mail servers, resource planning, and office automation systems. Level 2 (supervisory control local area network) includes the functions involved in monitoring and controlling the physical processes and the general deployment of systems such as HMIs, engineering workstations, and history logs. Level 1 (control network) includes the functions involved in sensing and manipulating physical processes. For example, receiving the information, processing the data, and triggering outputs, which are all done in PLCs. Level 0 (I/O network) consists of devices (sensors/actuators) that are directly connected to the physical process.

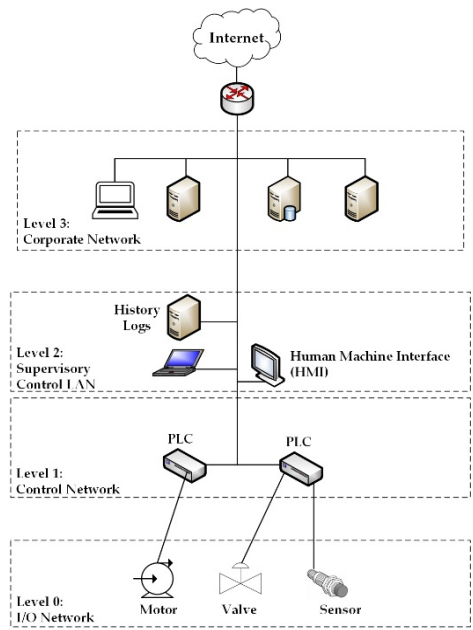

**Figure 1.** ICS reference model [4].

As shown in Figure 1, Level 3 is composed of the traditional IT infrastructure system (Internet access service, file transfer protocol server, Virtual Private Network (VPN) remote access, etc.). Levels 2, 1, and 0, represent a typical SCADA system which is composed of the following components:

- HMI: Used to observe the status of the system or to adjust the system parameters for processes control and management purposes.
- Engineering Workstation: Used by engineers for programming the control functions of the HMI.
- History Logs: Used to collect the data in real-time from the automation processes for current or later analysis.
- PLCs: Act as slave stations in the SCADA architecture. They are connected to sensors or actuators.

*2.2. The SCADA Communication Protocol*

There are several communication protocols developed for use in SCADA systems. These protocols define the standard message format for all inter-device communications on the network. One popular protocol, which is widely used in SCADA system environments, is Modbus protocol [5]. Modbus is an application-layer messaging protocol that provides the Client/Server communications between devices connected to an Ethernet network and offers services specified by function codes. The function codes tell the server what action to take. For example, a client can read the status of the discrete outputs or the values of digital inputs from the PLC; or it can read/write the data contents of a group of registers inside the PLC. Figure 2 illustrates an example of Modbus Client/Server communication.

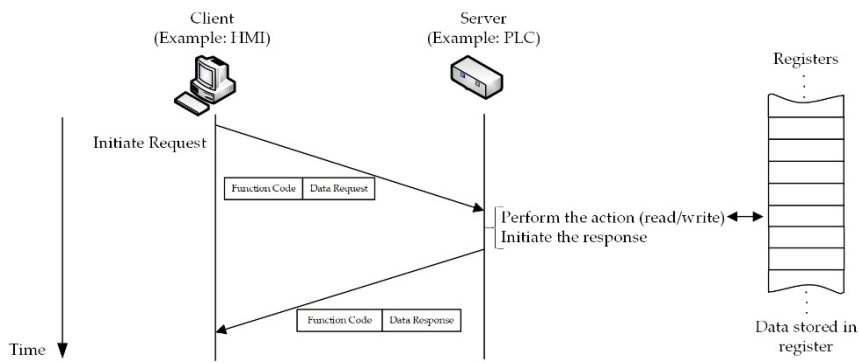

**Figure 2.** Modbus Client/Server communication example.

The Modbus register address type consists of four data reference types [5, 6] which are summarized in Table 1. The "xxxx" following a leading digit represents a four-digit address location in the user data memory.

**Table 1.** Data reference types [6, 7].

| Reference | Range | Description |
|---|---|---|
| 0xxxx | 00001-09999 | Read/Write Discrete Outputs or Coils. |
| 1xxxx | 10001-19999 | Read Discrete Inputs. |
| 3xxxx | 30001-39999 | Read Input Registers. |
| 4xxxx | 40001-49999 | Read/Write-Output or Holding Registers. |

*2.3. Related Works*

Cyber-attacks are continuously evolving and changing behavior to bypass the security mechanisms. Thus, the utilization of advanced security mechanisms is essential to identify and prevent new attacks. In this sense, the development of real testbeds enables the possibility of advance research.

Morris et al. [7] describe four datasets to be used for cybersecurity research. The datasets include network traffic, process control, and process measurement features from a set of attacks against testbeds which use Modbus application layer protocol. The authors argue that there are several datasets developed to train and validate IDS associated with traditional information technology systems, but in the SCADA security area, there is a lack of availability and access to SCADA network traffic. In our work, a new dataset with new types of attacks has been created. So, we are providing one more resource to be used by researchers to train, validate, and compare their results with other datasets once our dataset is available.

In order to investigate the security of the Modbus/TCP protocol, Miciolino et al. [8] explore a complex cyber-physical testbed, conceived for the control and monitoring of a water system. The analysis of the experimental results highlights the critical characteristics of the Modbus/TCP as a popular communication protocol in ICS environments. They conclude that by obtaining sufficient knowledge of the system, an attacker is able to change the commands of the actuators or the sensor readings in order to achieve its malicious objectives. Obtaining knowledge of the system is the first step in attacking a system. This attack is also known as reconnaissance attack. Hence, in our work, our ML models are trained to recognize this kind of attack.

In [9], Rosa et al. describe some practical cyber-attacks using an electricity grid testbed. This testbed consists of a hybrid environment of SCADA assets (e.g., PLCs, HMIs, process control servers) controlling an emulated power grid. The work explains their attacks and discusses some of the challenges faced by an attacker in implementing them. One of the attacks is the reconnaissance network attack. The authors argue that this kind of attack can be used not only to discover devices and types of services but also to perform fingerprinting and discover PLCs behind the gateways. Hence, in our work, advanced reconnaissance attacks are carried out, and ML algorithms are used to detect them.

In [10], Keliris et al. develop a process-aware supervised learning defense strategy that considers the operational behavior of an ICS to detect attacks in real-time. They use a benchmark chemical process and consider several categories of attack vectors on their hardware controllers. They use their trained SVM model to detect abnormalities in real-time and to distinguish between disturbances and malicious behavior as well. In our work, we use five ML algorithms to identify the abnormal behavior in real-time and evaluate their detection performance.

Tomin et al. [11] present a semi-automated method for online security assessment using ML techniques. They outline some experience obtained at the Melentiev Energy Systems Institute, Russia in developing ML-based approaches for detecting potentially dangerous states in power systems. Multiple ML algorithms are trained offline using resampling cross-validation method. Then, the best model among the ML algorithms is selected based on performance and is used online. They argue that the use of ML techniques provides reliable and robust solutions that can resolve the challenges in planning and operating future industrial systems with an acceptable level of security.

Cherdantseva et al. [12] review state of the art in cybersecurity risk assessment of SCADA systems. This review indicates that despite the popularity of the machine learning techniques, research groups in ICS security have reported the lack of standard datasets for training and testing machine learning algorithms. The lack of standard datasets has resulted in an inability to develop robust ML models to detect the anomalies in ICS. Using the testbed proposed in this paper, we built a new dataset for training and testing ML algorithms.

## 3. The SCADA System Testbed

In this section, we describe the configuration of our SCADA system testbed for cybersecurity research.

### 3.1. The Testbed Framework

The purpose of our testbed is to emulate real-world industrial systems as closely as possible without replicating an entire plant or assembly system [13]. The utilization of a testbed allows the possibility of carrying out real cyber-attacks. Our testbed is dedicated to controlling a water storage tank, which is a part of the process of water treatment and distribution. The components used in our testbed are commonly used in real SCADA environments. Figure 3 shows the SCADA testbed framework for our targeted application and Table 2 shows a brief description of the equipment used to build the testbed.

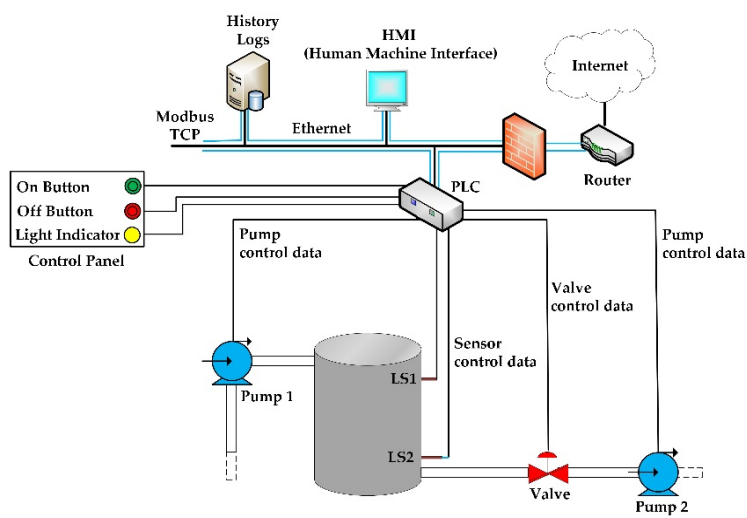

**Figure 3.** The testbed framework.

**Table 2.** Description of the devices used in the testbed.

| Devices | Descriptions |
| --- | --- |

| On Button: | Turns on the level control process of the water storage tank. |
|---|---|
| Off Button: | Turns off the level control process of the water storage tank. |
| Light Indicator: | Indicates whether the system is on or off. |
| Level Sensor 1 (LS1): | Monitors the maximum water level in the tank. When the water reaches the maximum level, the sensor sends a signal to PLC. |
| Level Sensor 2 (LS2): | Monitors the minimum water level in the tank. When the water reaches the minimum level, the sensor sends a signal to PLC. |
| Valve: | Controls the water level in the tank. When the water reaches the maximum level, the valve opens, and when the water reaches the minimal level, the valve closes. This logic is implemented in PLC using the ladder language. |
| Water Pump 1: | Fills up the water tank. |
| Water Pump 2: | Draws water from the tank when the valve is open. |
| PLC: | Controls the physical process. The logic of the water control system is in PLC, which receives signals from the input devices (buttons, sensors), executes the program, and sends signals to the output devices (water pumps and valve). |
| HMI: | Used by the administrator to monitor and control the water storage system in real-time. The administrator can also display the devices' state and interact with the system through this interface. |
| Data History: | Used to store logs and events of the SCADA system. |

As shown in Figure 3, the storage tank has two level sensors: Level Sensor 1 (LS1) and Level Sensor 2 (LS2) that monitor the water level in the tank. When the water reaches the maximum level defined in the system, the LS1 sends a signal to the PLC. The PLC turns off Water Pump 1 used to fill up the tank, opens the valve, and turns on Water Pump 2 to draw the water from the tank. When the water reaches the minimal level defined in the system, LS2 sends a signal to the PLC, which closes the valve, turns off Water Pump 2, and turns on Water Pump 1 to fill up the tank. This process starts over when the water level reaches LS1. SCADA system gets data from the PLC using the Modbus communication protocol and displays them to the system operator through the HMI interface.

There are other ICS protocols which could be used instead of Modbus in our testbed. For example, DNP3 is an ICS protocol that provides some security mechanisms [14,15]. However, in a recent research, Li et al. [16] reported that they found 17,546 devices connected to the Internet using the Modbus protocol spread all over the world. They have not counted the number of equipment not directly connected to the internet. Although there are other ICS protocols, many industries still use SCADA systems with Modbus protocol because their equipment does not support other protocols. In this case, solutions to detect attacks can be cheaper instead of other solution like changing the devices.

PLC Schneider model M241CE40 is used in our testbed to control the process of the water storage tank. The logic programming of the PLC is done using the LADDER programming language. The LADDER language is not covered in this paper; however, more information can be found in [17, 18]. The sensors described in Table 2 are connected to the digital inputs of the PLC. The pumps and valves are connected to the output of the PLC.

## 4. Machine Learning Algorithms and Performance Measurements

In this section, we describe the ML algorithms used in our work as well as the measurements used to evaluate their performances.

*4.1. Machine Learning Algorithms*

The ML algorithms can be classified as supervised, unsupervised, and semi-supervised. Each class has its characteristics and applicability. The discussion of all algorithms is out of the scope of this paper. However, we refer the reader to [19] [20] for detailed technical discussions of these algorithms. In this paper, we use traditional ML algorithms to detect the attacks. Our target is to build supervised machine learning models, and we have chosen the followings algorithms for attack detection and classification:

1. Logistic Regression [20].
2. Random Forest [21].
3. Naïve Bayes [22].
4. Support Vector Machine (SVM) [23].
5. KNN [24].

The performance of these algorithms is discussed in Section 6.

*4.2. Performance Measurements*

Traditionally, the performance of ML algorithms is measured by metrics which are derived from the confusion matrix [25]. Table 3 shows the confusion matrix in the IDS context.

Table 3. Confusion matrix in IDS context.

| Data class | Classified as normal | Classified as abnormal |
|---|---|---|
| Normal | True Negative (TN) | False Negative (FN) |
| Abnormal | False Positive (FP) | True Positive (TP) |

In the IDS context, the following parameters are used to compose the confusion matrix:

- TN: Represents the number of normal flows correctly classified as normal (e.g., normal traffic);
- TP: Represents the number of abnormal flows (attacks) correctly classified as abnormal (e.g., attack traffic);
- FP: Represent the number of normal flows incorrectly classified as abnormal;
- FN: Represents the number of abnormal flows incorrectly classified as normal;

Next, we present several evaluation metrics and their respective formulas which are derived from the confusion matrix parameters:

- Accuracy: It is the percentage of correctly predicted flows considering the total number of predictions:

$$\text{Accuracy \%} = \frac{TP + TN}{TP + TN + FP + FN} \times 100 \quad (1)$$

- False Alarm Rate (FAR): This represents the percentage of the normal flows misclassified as abnormal flows (attack) by the model:

$$\text{FAR \%} = \frac{FP}{FP + TN} \times 100 \quad (2)$$

- UN-Detection Rate (UND): It is the fraction of the abnormal flows (attack) which are misclassified as normal flows by the model:

$$\text{UND \%} = \frac{FN}{FN + TP} \times 100 \quad (3)$$

Accuracy (Eq. 1) is the most frequently used metric for evaluating the performance of learning models in classification problems. However, this metric is not very reliable for evaluating the ML performance in the scenarios with imbalanced classes [26]. In this case, one class is dominant in number, and it has more samples relatively compared to another class. For example, in IDS scenarios, the proportion of normal flows to attack flows is very high in any realistic dataset. That is, the number of samples in the dataset which represent the normal flows is enormous compared to the number of samples which represent the attack flows. This problem is prevalent in scenarios where anomaly detection is crucial like fraudulent transactions in banks, identification of rare diseases, and in the identification of cyber-attacks in critical infrastructures. New metrics have been developed to avoid a biased analysis [27]. So, in addition to the accuracy, we also used the FAR and UND metrics.

## 5. Attack Scenarios, Features Selection, and Evaluation Scenarios

In this section, we describe the attacks carried out in our testbed and the features used to build our dataset. This dataset will be used in Section 6 for training and testing the ML algorithms.

*5.1. Attack Scenarios*

Network attacks on SCADA systems can be divided into three categories: Reconnaissance, Command Injection, and Denial of Service (DoS) [7]. Our focus in this paper is on the reconnaissance attacks where the network is scanned for possible vulnerabilities to be used for later attacks. Reconnaissance attack is the first stage of any attack on a networking system. In this stage, hackers use scan tools to inspect the topology of the victim network and identify the devices in the network as well as their vulnerabilities. Figure 4 shows our testbed attack scenario where the dashed rectangles highlight the vulnerable spots and possible attack targets in the system.

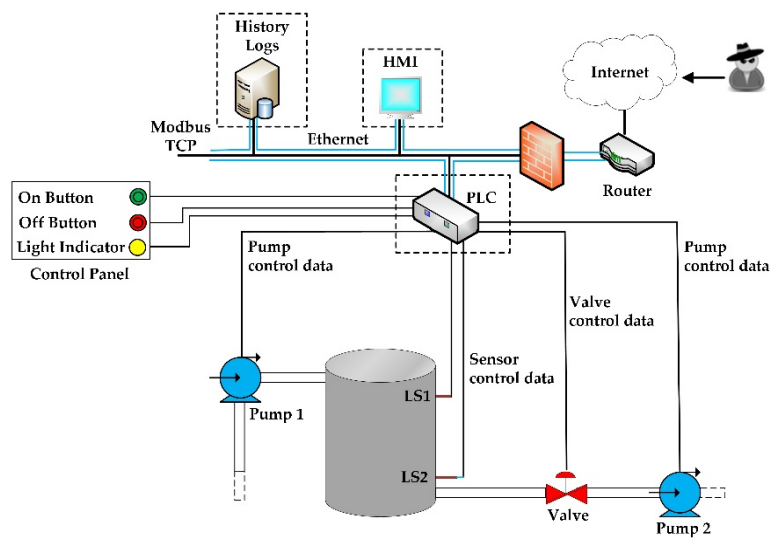

**Figure 4.** Attack Scenario.

Some reconnaissance attacks can be easily detected. For example, there are scanning tools which send a large number of packets per second under Modbus/TCP to the targeted device and wait for acknowledgment of the packets from them. If a response is received, the host (i.e., the device) is active. This attack generates a considerable variation in the traffic behavior which can be easily detected by the traditional IDS or even the traditional firewall or rule-based mechanisms. Figure 5 shows an example of the traffic behavior when a scanning tool was used in our testbed:

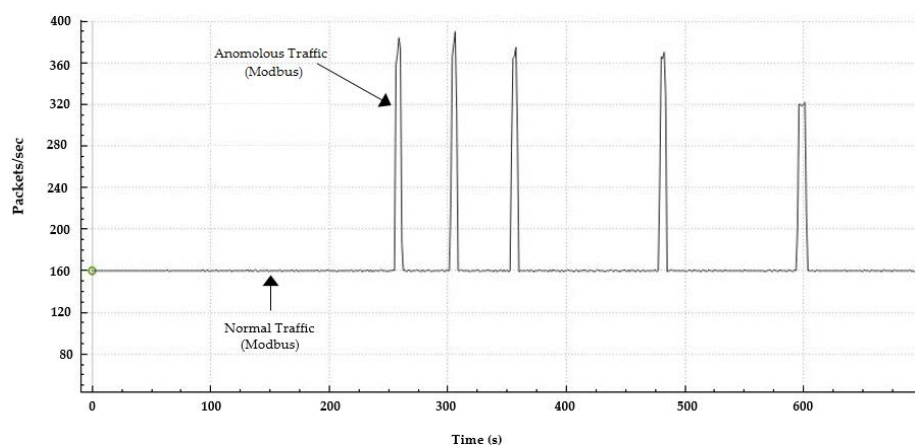

**Figure 5.** Network traffic behavior under easy to detect attacks.

On the other hand, there are some sophisticated reconnaissance attacks which are more difficult to detect. For example, some exploits can be used to map the network, and that results in an attack behavior almost similar to normal traffic. Figure 6 illustrates the network traffic behavior during such exploit attacks. As can be seen, the change in the traffic behavior is negligible under the attack. Thus, it is difficult to detect the attack. The use of rule-based mechanisms would fail because the signature of the Modbus and TCP traffic do not change, and the language used to express the detection rules may not be expressive enough. On the other hand, the use of ML can improve the detection rate as ML algorithms can be trained to detect these attack scenarios.

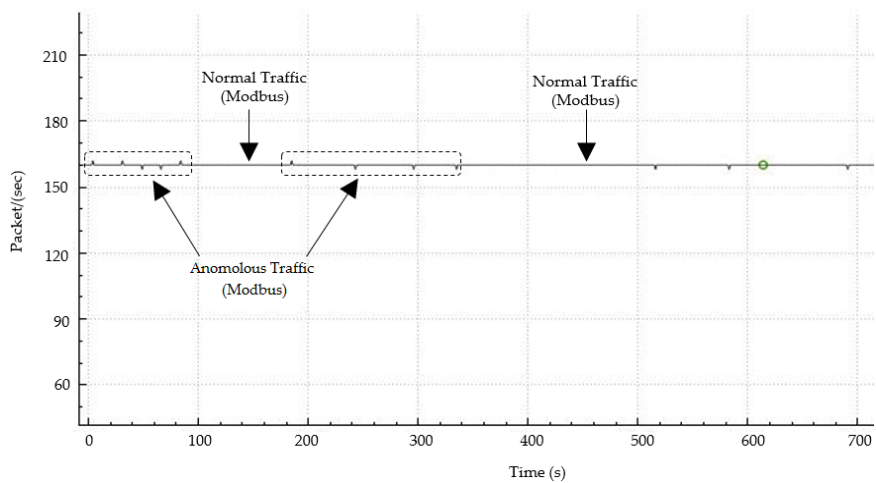

**Figure 6.** Network traffic behavior under difficult to detect attacks.

We conducted the following reconnaissance and exploit attacks specific to ICS environment described in Table 4. Details of the commands used to perform the attacks can be found in [28] and [29]. During the attacks, the network traffic was captured to be analyzed. We used the following tools to analyze the captured traffic: Wireshark [30], and Argus [31]. The captured traffic included unencrypted control information of the devices (valve, pumps, sensors) as well as information regarding their type (function codes, type of data). Table 5 presents statistical information about the captured traffic.

**Table 4.** Reconnaissance attacks carried out against our testbed [29,30].

| Attack Name | Attack Description |
|---|---|
| Port Scanner [29] | This attack is used to identify common SCADA protocols on the network. Using Nmap tool, packets are sent to the target at intervals which vary from 1 to 3 seconds. The TCP connection is not fully established so that the attack is difficult to detect by rules. |
| Adress Scan Attack [29] | This attack is used to scan network addresses and identify the Modbus server address. Each system has only one Modbus server and disabling this device would collapse the whole SCADA system. Thus, this attack tries to find the unique address of the Modbus server so that it can be used for further attacks. |
| Device Identification Attack [29] | This attack is used to enumerate the SCADA Modbus slave IDs on the network and to collect additional information such as vendor and firmware from the first slave ID found. |
| Device Identification Attack (Aggressive Mode) [29] | This attack is similar to the previous attack. However, the scanning uses an aggressive mode, which means that the additional information about all slave IDs found in the system us collected. |
| Exploit [30] | Exploit used to read the coils values of the SCADA devices. The coils represent the ON/OFF status of the devices controlled by the PLC, such as motors, valves, and sensors [29]. |

**Table 5.** Statistical information on the captured traffic.

| Measurement | Value |
|---|---|
| Duration of capture (hours) | 25 |
| Dataset length (GB) | 1.27 |
| Number of observations | 7,049,989 |
| Average data rate (kbits/s) | 419 |
| Average packet size (bytes) | 76.75 |
| Percentage of scanner attack | $3 \times 10^{-4}$ |
| Percentage of address scan attack | $75 \times 10^{-4}$ |
| Percentage of device identification attack | $1 \times 10^{-4}$ |
| Percentage of device identification attack (agressive mode) | 4.93 |
| Percentage of exploit attack | 1.13 |
| Percentage of all attacks (total) | 6.07 |
| Percentage of normal traffic | 93.93 |

*5.2. Features Selection*

Once the network traffic is captured, the next step is to select potential features which can distinguish the anomalous traffic from the normal traffic. The authors in [19] have selected 12 useful features for ML-based network security monitoring in the ICS networks. In [32], the authors study the potential features presented in [19]. In our work, we analyzed the variation of the features during the normal and attack traffic, and we analyzed those features that do not vary during the normal and attack traffic. Based on these prior works and in our studies, Table 6 shows the features selected for our dataset.

Table 6. Features selected to compose the dataset.

| Features | Descriptions |
|---|---|
| Total Packets (TotPkts) | Total transaction packet count |
| Total Bytes (TotBytes) | Total transaction bytes |
| Source packets (SrcPkts) | Source/Destination packet count |
| Destination Packets (DstPkts) | Destination/Source packet count |
| Source Bytes (SrcBytes) | Source/Destination transaction bytes |
| Source Port (Sport) | Port number of the source |

*5.2. Evaluation Scenario*

After defining the dataset, the features are extracted as discussed in the previous subsection. Then, the data is labeled either as normal traffic or attack traffic. Following that, the dataset is split into training and test datasets. The training dataset is composed of 80% of the total data, and it is used to train our ML model. The test dataset consists of the remaining 20% of the data, and it is used to evaluate the performance of our trained ML model. We call this training and test phase as "offline evaluation," because the ML models are trained and tested offline. Figure 7 shows our evaluation scenario.

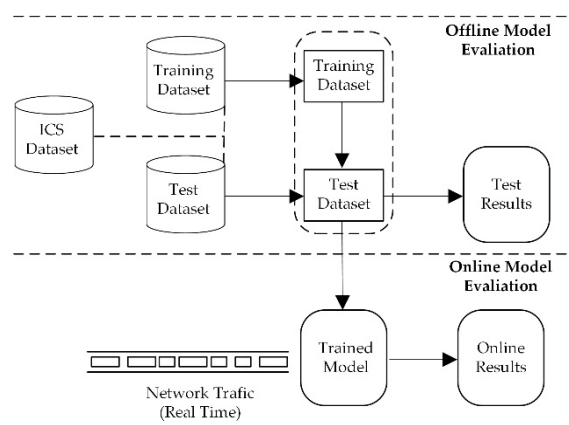

**Figure 7.** Model Evaluation.

After training and testing, the trained ML models are created and deployed in the network. Then, their performance is analyzed using real network traffic. This phase is called "online evaluation." We compare the results obtained from the two phases (offline and online). This is described next.

**6. Numerical Results**

In this section, we present the numerical results of the attacks described in Subsection 5.1. Figure 8 shows the results about the accuracy of the ML algorithms used.

The accuracy represents the total number of correct predictions divided by the total number of samples (Eq. 1). As shown in Figure 8, considering the offline evaluations, Decision Tree and KNN have the best accuracy (100%) compared to other ML models. However, the difference in the accuracy is small among all trained models. In other words, all chosen ML algorithms performed well in terms of accuracy during the offline phase. During the online phase, Decision Tree, Random Forest, Naïve Bayes and Logistic Regression have a small difference, hence, the performance of these algorithms, in both phases (offline and online), are similar. The same does not happen with the KNN model. There is a significant difference between the online and offline phase which indicates that in practice the KNN does not provide good accuracy.

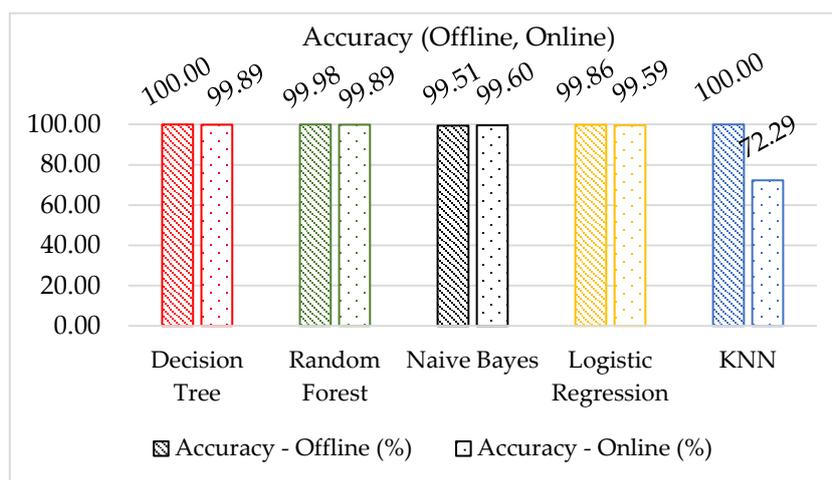

**Figure 8.** Accuracy results.

As shown in Table 5, our dataset is unbalanced. Therefore, accuracy is not the ideal measure to evaluate the performance [33]. Other metrics are needed to compare the performance of the ML algorithms. Figure 9 shows the False Alarm Rate results. The FAR metric is the percentage of the regular traffic which has been misclassified as anomalous by the model (Eq. 2).

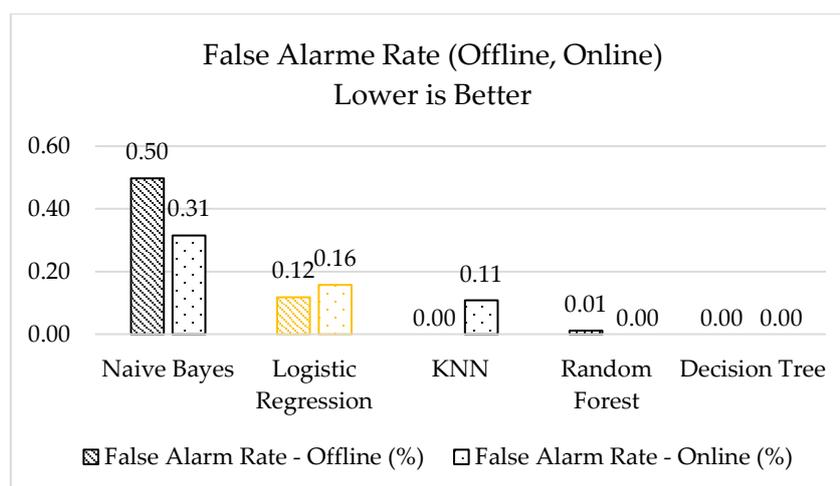

**Figure 9.** False alarm rate results.

Regarding the offline and online evaluations, as shown in Figure 9, Random Forest and Decision Tree models have the best performance followed by the KNN model. These three models have the lowest false alarm percentages followed by the Logistical Regression and Naïve Bayes. These lowest percentages mean that Random Forest, Decision Tree, and KNN perform better in detecting normal traffic. In our dataset, normal traffic is the dominant traffic; therefore, it is expected to have a low FAR value. This low FAR value can be due to the model bias toward estimating the normal traffic perfectly, which is common in unbalanced datasets. Further, the clustering done in the Random Forest, Decision Tree and KNN models can be helpful, especially when dealing with two types of data having different network features.

Figure 10 shows the results of the un-detection rate metric. The UND (Eq. 4) represents the percentage of the traffic which is an anomaly but is misclassified as normal (the opposite of the FAR). The traffic represented by this metric is more critical than the traffic represented by the FAR metric because, in this case, an attack can happen without being detected. Further, in our unbalanced dataset, the models are biased toward normal traffic and this metric would show how biased the models are.

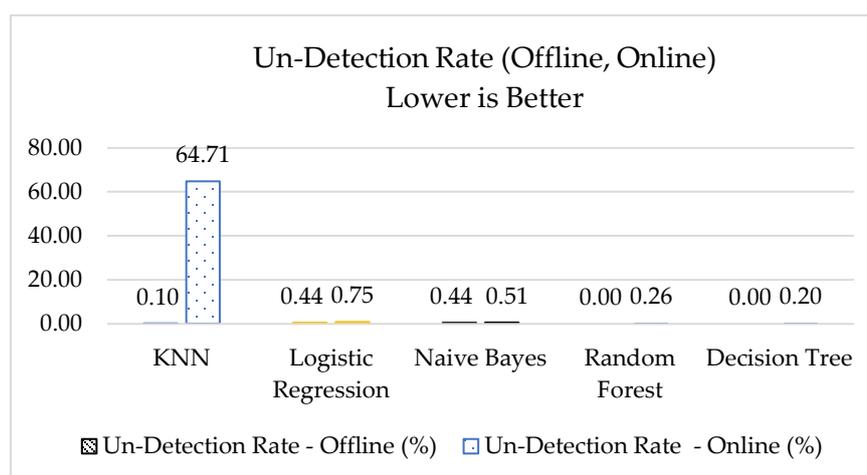

**Figure 10.** Un-Detected rate results.

As shown in Figure 10, considering the offline performance results, the percentage of the UND is small for Naïve Bayes, Logistic Regression, and KNN models, and zero for Decision Tree and Random Forest models. That is, all algorithm shows excellent performance on this critical metric. However, considering the online performances, KNN model has the worst performance, much different from the offline evaluation. The same does not happen to the other models, where their online performances are very close to the offline performance. This excellent performance shows that our selected features in this work are great as they could detect attacks even in an unbalanced dataset.

## 7. Conclusion

This paper presented the development of a SCADA system testbed to be used in cybersecurity research. The testbed is dedicated to controlling a water storage tank which is one of the several stages in the process of water treatment and distribution. The testbed was used to analyze the effects of the attacks on SCADA systems. Using the network traffic, a new dataset was developed for use by researchers to train machine learning algorithms as well as validating and comparing their results with other available datasets.

Five reconnaissance attacks specific to ICS environment were conducted against the testbed. During the attacks, the network traffic with information about the devices (valves, pumps, sensors) was captured. Using Argus and Wireshark network tools, features were extracted to build a dataset for training and testing machine learning algorithms.

Once the dataset was generated, five traditional machine learning algorithms were used to detect the attacks: Random Forest, Decision Tree, Logistic Regression, Naïve Bayes and KNN. These algorithms were evaluated in two phases: during the training and test of the machine learning models (offline), and during the deployment of these models in the network (online). The performance obtained during the online phase was compared to the performance obtained during the offline phase

Three metrics were used to evaluate the performance of the used algorithms: accuracy, FAR, and UND. Regarding the accuracy metric, in the offline phase, all ML algorithms showed an excellent performance. In the online phase, almost all the algorithms had their performance very close to the offline results. The KNN algorithm was the only one which did not perform well. Moreover, considering an unbalanced dataset and analyzing the FAR and UND metrics, we concluded that Random Forest and Decision Tree models have better performance compared to others in both phases.

Results show the feasibility of detecting reconnaissance attacks in ICS environments. Our future plans include generating more attacks and checking for the models' feasibility and performance in different environments. Moreover, experiments using unsupervised algorithms will be done.

**Acknowledgments:** This work has been supported under the grant ID NPRP 10-901-2-370 funded by the Qatar National Research Fund (QNRF) and grant#2017/01055-4 São Paulo Research Foundation (FAPESP). The statements made herein are solely the responsibility of the authors. The authors would like to thank the Instituto Federal de Educação, Ciência e Tecnologia de São Paulo (IFSP), Washington University in Saint Louis, and Qatar University.

**Author Contributions:** Marcio Andrey Teixeira built the testbed and performed the experiments. Tara Salman and Maede Zolanvari assisted with revisions and improvements. The work was done under the supervision and guidance of Professors Jain, Meskin, and Samaka, who also formulated the problem.

**Conflicts of Interest:** The authors declare no conflicts of interest.